\documentclass[11pt]{article}
\PassOptionsToPackage{authoryear}{natbib}
\usepackage[preprint,nonatbib]{neurips_2024}
\usepackage{hyperref} 
\usepackage{authblk} 
\usepackage{booktabs}
\usepackage{graphicx}
\usepackage{amsfonts}
\usepackage{multicol}
\usepackage{tabularx}
\usepackage{enumitem,amsmath,bm,amssymb}
\usepackage{pifont,makecell}
\usepackage{tikz}
\usepackage{CJKutf8}
\newcommand*\circled[1]{\tikz[baseline=(char.base)]{
		\node[shape=circle,draw,inner sep=0.5pt] (char) {#1};}}
\usepackage{multirow}
\usepackage{cite}
\usepackage{comment}

\title{CosyVoice 2: Scalable Streaming Speech Synthesis with Large Language Models}

\author{
  \bf Zhihao Du, Yuxuan Wang, Qian Chen, Xian Shi, Xiang Lv, Tianyu Zhao, Zhifu Gao\\
  \bf Yexin Yang, Changfeng Gao, Hui Wang, Fan Yu, Huadai Liu, Zhengyan Sheng \\
  \bf Yue Gu, Chong Deng, Wen Wang, Shiliang Zhang, Zhijie Yan,  Jingren Zhou\thanks{The code and pre-trained models are released at: \url{https://github.com/FunAudioLLM/CosyVoice}}
}

\affil{Alibaba Group, China}
\affil{\texttt{\{neo.dzh,sly.zsl\}@alibaba-inc.com}}
\date{}
\begin{document}

\maketitle

\begin{abstract}
In our previous work, we introduced CosyVoice, a multilingual speech synthesis model based on supervised discrete speech tokens. By employing progressive semantic decoding with two popular generative models, language models (LMs) and Flow Matching, CosyVoice demonstrated high prosody naturalness, content consistency, and speaker similarity in speech in-context learning. Recently, significant progress has been made in multi-modal large language models (LLMs), where the response latency and real-time factor of speech synthesis play a crucial role in the interactive experience. Therefore, in this report, we present an improved streaming speech synthesis model, CosyVoice 2, which incorporates comprehensive and systematic optimizations.
Specifically, we introduce finite-scalar quantization to improve the codebook utilization of speech tokens. 
For the text-speech LM, we streamline the model architecture to allow direct use of a pre-trained LLM as the backbone.
In addition, we develop a chunk-aware causal flow matching model to support various synthesis scenarios, enabling both streaming and non-streaming synthesis within a single model.  
By training on a large-scale multilingual dataset, CosyVoice 2 achieves human-parity naturalness, minimal response latency, and virtually lossless synthesis quality in the streaming mode. We invite readers to listen to the demos at \url{https://funaudiollm.github.io/cosyvoice2}.
\end{abstract}

\section{Introduction}

In recent years, neural text-to-speech (TTS) synthesis models have garnered significant attention for surpassing traditional concatenative and statistical parametric methods \cite{DBLP:conf/interspeech/WangSSWWJYXCBLA17, DBLP:conf/icassp/ShenPWSJYCZWRSA18, DBLP:journals/corr/abs-1710-07654, DBLP:conf/iclr/PingPC19, DBLP:conf/nips/RenRTQZZL19, DBLP:conf/aaai/Li0LZL19, DBLP:conf/iclr/0006H0QZZL21}. These models have achieved high fidelity and naturalness on pre-defined specific speakers. Recent studies show that zero-shot TTS models are able to synthesize speech for any speaker by imitating the timbre, prosody and style of a reference speech \cite{DBLP:journals/corr/abs-2301-02111}. Beyond their in-context learning (ICL) capability, zero-shot TTS models benefit from large-scale training data, achieving synthesis quality and naturalness nearly indistinguishable from human speech.

Recent zero-shot TTS models can be broadly divided into three categories: codec language models, feature diffusion models and their hybrid systems. Codec language models utilize a speech codec model to extract discrete speech representation \cite{DBLP:journals/taslp/ZeghidourLOST22,DBLP:journals/tmlr/DefossezCSA23,DBLP:conf/icassp/DuZHZ24} and employ an autoregressive \cite{DBLP:journals/corr/abs-2301-02111,DBLP:journals/tacl/KharitonovVBMGP23,DBLP:journals/corr/abs-2401-07333,DBLP:journals/corr/abs-2401-14321,DBLP:journals/corr/abs-2404-03204,DBLP:journals/corr/abs-2406-05370,DBLP:journals/corr/abs-2406-07855} or masked \cite{DBLP:journals/corr/abs-2409-00750} language model to predict the speech tokens, which are then synthesized to waveforms via codec vocoders \cite{DBLP:conf/asru/OkamotoYOTK23,DBLP:conf/iclr/Siuzdak24}. Continuous speech representations are also explored in \cite{DBLP:journals/corr/abs-2407-08551}. Language model-based TTS can generate varied and prosody-consistent speech via autoregressive sampling.

Inspired by advances in image generation, denoising diffusion \cite{DBLP:conf/nips/HoJA20,DBLP:conf/iclr/0011SKKEP21} and flow matching models \cite{DBLP:conf/iclr/LipmanCBNL23} have been introduced into non-autoregressive (NAR) speech synthesis. Early diffusion-based TTS models required duration prediction for each text (phone) to address the length disparity between text and speech features \cite{DBLP:conf/nips/LeVSKSMWMAMH23,DBLP:conf/icml/JuWS0XYLLST000024,DBLP:conf/icassp/GuoDM0024,DBLP:conf/icassp/MehtaTBSH24}. However, this rigid alignment can affect naturalness, resulting in flat prosody. To mitigate this issue, cross-attention and Diffusion Transformers (DiT) have been introduced into NAR TTS models \cite{DBLP:conf/asru/GaoMZC23,DBLP:journals/corr/abs-2406-11427}. Recent research indicates simpler approaches for text-speech alignment in NAR TTS models, such as E2 TTS \cite{DBLP:journals/corr/abs-2406-18009}, F5-TTS \cite{DBLP:journals/corr/abs-2410-06885} and Seed-TTS \cite{DBLP:journals/corr/abs-2406-02430}. In these models, input text is padded with special tokens to match the total speech length which is either automatically predicted by the utterance duration prediction module or specified by the user in advance. Since NAR TTS models are not constrained by codec vocoders, they can achieve superior speech quality.

Hybrid systems combine the text-to-codec language model and codec-to-feature diffusion model \cite{DBLP:journals/corr/abs-2406-02430,cosyvoice,DBLP:journals/corr/abs-2409-03283}. The language model addresses the alignment between text and speech as well as the utterance duration prediction, while the codec-to-feature diffusion model synthesizes speech features (Mel spectrum) based on the generated codec and other conditions. By leveraging the strengths of both generative models, hybrid systems achieve high diversity, prosody consistency and speech quality.

Despite the success of recent zero-shot TTS models, they generally operate in non-streaming (offline) mode, which involves complete input text and requires synthesizing the entire utterance before returning the waveform. This results in high latency, negatively impacting user experience in applications like voice chat \cite{hurst2024gpt,DBLP:conf/emnlp/ZhangLZZWZQ23}.
To address this issue, streaming synthesis has been explored for language model-based zero-shot TTS models \cite{DBLP:journals/corr/abs-2406-02897,DBLP:journals/corr/abs-2410-00767,DBLP:journals/corr/abs-2402-08093,DBLP:conf/icassp/DekelSFHKH24}, but diffusion-based TTS models and hybrid systems lack well-established streaming solutions. 

Building on the success of CosyVoice \cite{cosyvoice}, we introduce CosyVoice 2, a streaming zero-shot TTS model with improved prosody naturalness, content consistency, and speaker similarity. Our contributions include:
\begin{itemize}[leftmargin=*]
\item Unifying streaming and non-streaming synthesis in a single framework and proposing the unified text-speech language model and chunk-aware causal flow matching model, leading to lossless streaming synthesis compared to offline mode.
\item Simplifying the LM architecture by removing the text encoder and speaker embedding, allowing pre-trained textual large language models (LLMs) to serve as the backbone, enhancing context understanding.
\item Replacing vector quantization (VQ) in the speech tokenizer with finite scalar quantization (FSQ), improving codebook utilization and capturing more speech information.
\item Upgrading the instructed TTS capacity to support more instructions, including emotion, accent, role style, and fine-grained control. In CosyVoice 2, the instruction and zero-shot capacity are integrated into a single model, enabling more versatile and vivid synthesis.
\end{itemize}
Through the above systemic modification and optimization, CosyVoice 2 achieves human-parity synthesis quality and is nearly lossless in streaming mode. The unified framework loosens deployment requirements, enabling a single model to support both streaming and non-streaming synthesis. The upgraded instructed TTS capacity provides a more powerful and  easier approach for users to generate various speeches. In addition, the chunk-aware flow matching design can also be applied to NAR TTS models, which suggests the potential for streaming NAR models.

\section{CosyVoice 2}

\begin{figure*}[thb]
	\centering
	\includegraphics[width=\linewidth]{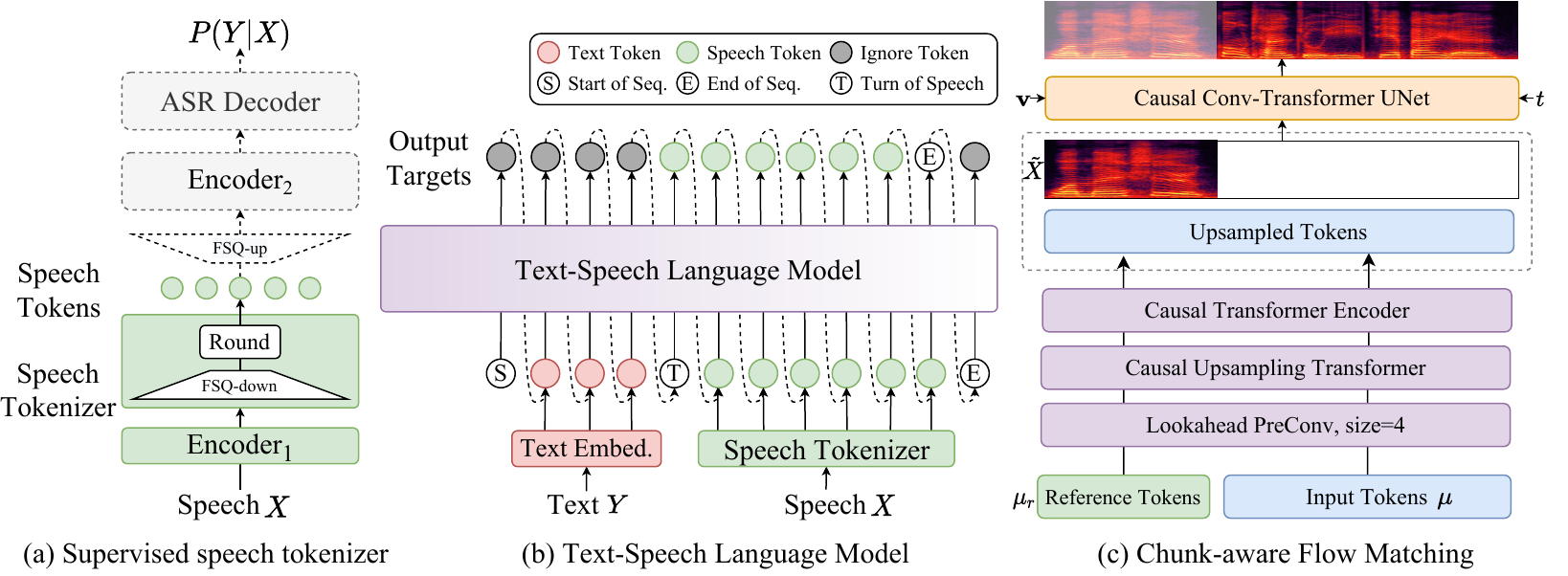}
	\vspace{-0.6cm}
	\caption{An overview of CosyVoice 2. (a) demonstrates the supervised speech tokenizer, where dashed modules are only used at the training stage. (b) is a unified text-speech language model for streaming and non-streaming synthesis. Dashed lines indicate the autoregressive decoding at the inference stage. (c) illustrates the causal flow matching model conditioning on a speaker embedding $\mathbf{v}$, semantic tokens $\mu$, masked speech features $\tilde{X}$ and intermediate state $X_t$ at timestep $t$ on the probabilistic density path.}
	\label{fig:overall}
\end{figure*}

CosyVoice 2 builds on the similar design philosophy of its predecessor \cite{cosyvoice} by separating the semantic and acoustic information of speech signals and modeling them independently. The speech generation process is redefined as a gradual semantic decoding procedure, where conditional information is progressively incorporated. Specifically, the text-speech language model (LM) focuses solely on semantic information, decoding high-level text tokens into supervised semantic speech tokens. In the Flow Matching model, acoustic details, such as timbre, are introduced through speaker embeddings and reference speech, converting speech tokens into the Mel spectrum for a given speaker. Finally, a pre-trained vocoder model reinstates the phases, transforming the Mel spectrum back into the original audio signal. The following sections will introduce the details of CosyVoice 2 and the modifications for streaming synthesis from five respects: text tokenizer, supervised semantic speech tokenizer, unified text-speech LM for streaming/non-streaming synthesis and chunk-aware Flow Matching model. Figure \ref{fig:overall} provides an overview of CosyVoice 2.

\subsection{Text Tokenizer}
CosyVoice 2 uses the raw text as input directly, which is tokenized using a BPE-based text tokenizer. This eliminates the need for a frontend model that obtains phonemes via the grapheme-to-phoneme (g2p) transformation. This approach not only simplifies the data preprocessing workflow but also enables the model to learn the pronunciations of words within various contexts in an end-to-end manner. Unlike the tokenizers commonly used in textual LLMs, CosyVoice 2 masks out the one-to-many tokens. This prevents the pronunciation of a token from becoming excessively long and reduces corner cases caused by data sparsity. Specifically, if a BPE token encodes more than one Chinese character, it will be masked out, and each character will be encoded separately during the tokenization process. Other languages, such as English, Japanese, and Korean, are not subject to special handling.

\subsection{Supervised Semantic Speech Tokenizer}
As shown in Figure \ref{fig:overall} (a), we insert the finite scalar quantization (FSQ) module \cite{DBLP:conf/iclr/MentzerMAT24} into the encoder of SenseVoice-Large ASR model \cite{funaduiollm}. At the training stage, the input speech $X$ goes through the $\mathrm{Encoder}_1$ to obtain the intermediate representations, where $\mathrm{Encoder}_1$ consists of six Transformer blocks with the rotary positional embedding \cite{DBLP:journals/ijon/SuALPBL24}. Then, the intermediate representations are fed into the FSQ module for quantization, and the quantized representations are passed through the rest of SenseVoice-Large modules, including $\mathrm{Encoder}_2$ and $\mathrm{ASR\ Decoder}$, to predict the posterior probabilities of corresponding text tokens. 

In the FSQ module, the intermediate representations $H$ are firstly projected into a $D$-dimensional low-rank space, and the values of each dimension are quantized into $[-K,K]$ with the bounded round operation $\mathrm{ROUND}$. Then, the quantized low-rank representations $\bar{H}$ are projected into the original dimension $\tilde{H}$ for the following modules:
\begin{equation}
\begin{split}
	\bar{H} &= \mathrm{ROUND}(\mathrm{Proj}_{down}(H)) \\
	\hat{H} &= \mathrm{Proj}_{up}(\bar{H})
\end{split}
\end{equation}
At the training stage, the straight-through estimation is used to approximate the gradients of FSQ module and $\mathrm{Encoder}_1$.
The speech token $\mu_i$ can be obtained by calculating the index of quantized low-rank representation $\bar{h}_i$ in the $(2K+1)$-ary system:
\begin{equation}\label{eq:codec}
	\mu_i = \sum_{j=0}^{D-1}{\bar{h}_{i,j}(2K+1)^{j}}
\end{equation}
The $\mathrm{Encoder}_1$, low-rank projector of FSQ module, bounded round operation and index calculation form the speech tokenizer for CosyVoice 2. Our speech tokenizer works at a token rate of 25 Hz, i.e., 25 speech tokens per second.

\begin{figure*}[thb]
	\centering
	\includegraphics[width=0.6\linewidth]{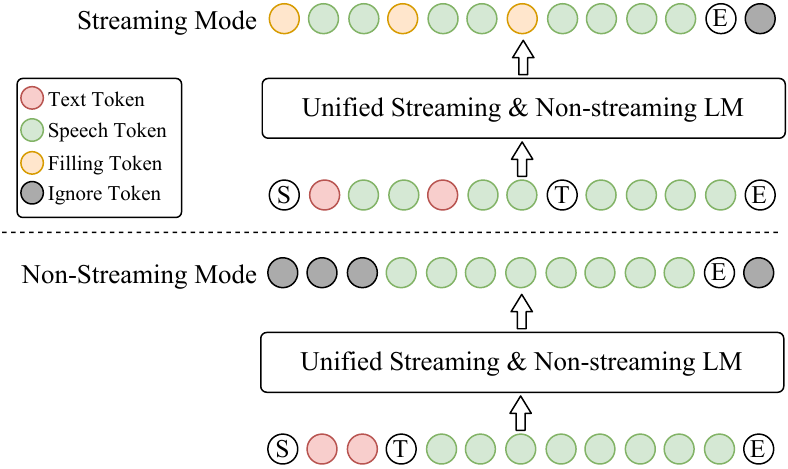}
	\caption{A diagram of the unified text-speech language model for streaming and non-streaming synthesis in CosyVoice 2.}
	\label{fig:ULM}
\end{figure*}
\subsection{Unified Text-Speech Language Model}
In CosyVoice 2, the pre-trained textual LLM, Qwen2.5-0.5B \cite{qwen2.5}, is used as the text-speech language model to generate the speech tokens autoregressively with the input text as a prompt.
Similar to other LMs, the text-speech LM is also trained in a next-token-prediction scheme as shown in Figure \ref{fig:overall} (b). 
Different from the previous CosyVoice, we remove the speaker embedding to avoid information leaking. More importantly, we find that such utterance-level vector contains not only speaker identify but also language and paralanguage information, which harms the prosody naturalness and cross-lingual capability of the text-speech LM.
Besides, we also abandon the text encoder of the previous CosyVoice, since we find that the Qwen2.5-0.5B model is powerful enough to align the text and speech tokens, and the text encoder is no longer needed.

Benefiting from the simplicity of text-speech LM, we can build a unified model for both streaming and non-streaming synthesis. Here, ``streaming mode'' means the input text is received in a continuous flow rather than being known as a complete sentence in advance. In CosyVoice 2, the difference between streaming and non-streaming modes is only the way of sequence construction for LM:
\begin{itemize}[leftmargin=*]
\item For the \textbf{Non-Streaming} mode, the ``start of sequence'' \circled{S}, all text tokens, ``turn of speech'' token \circled{T}, all speech tokens and the ``end of sequence'' \circled{E} are concatenated sequentially as shown in the bottom of Figure \ref{fig:ULM}. Ignore token means that their losses are ignored while minimizing the cross-entropy objective function.
\item For the \textbf{Streaming} mode, we mix up the text and speech tokens in a pre-defined ratio of $N$:$M$, i.e. every $N$ text tokens are followed by $M$ speech tokens seen in the top of Figure \ref{fig:ULM}. If the next token is a text token, the model is expected to predict a filling token (rather than the text token), which indicates that the next $N$ text tokens should be concatenated at the inference stage. Once the text tokens are ran out of, the ``turn of speech'' token \circled{T} and the remaining speech tokens are concatenated sequentially, forming the hybrid text-speech token sequence in the streaming mode. In our experiments, $N$ and $M$ are set to 5 and 15, respectively.
\end{itemize}
By training the text-speech LM on the above two sequences simultaneously, we can perform streaming and non-streaming speech generation within a single unified model. 
In real-life scenarios, such as speaker fine-tuning (SFT) and in-context learning (ICL), the inference sequence differs as follows:
\begin{itemize}[leftmargin=*]
\item \textbf{ICL, Non-Streaming:} In ICL, the LM requires prompt text and speech tokens from the reference audio to imitate the accent, prosody, emotion and style. In the non-streaming mode, the prompt and to-synthesize text tokens are concatenated as the whole entity, and the prompt speech tokens are treated as the pre-generated results and are fixed: ``\circled{\textbf{S}}, \textbf{prompt\_text}, \textbf{text}, \circled{\textbf{T}}, \textbf{prompt\_speech}". The autoregressive generation of LM is started from such sequence until the ``End of sequence'' token \circled{E} is detected.

\item \textbf{ICL, Streaming:} In this scenario, we assume the to-generate text is already known and the speech tokens should be generated in a streaming manner. Similarly, we treat the prompt and to-generate text as a whole entity. Then, we mix it up with the prompt speech tokens on the ratio of $N$:$M$: ``\circled{\textbf{S}}, \textbf{mixed\_text\_speech}, \circled{\textbf{T}}, \textbf{remaining\_speech}''. If the length of text is greater than that of prompt speech tokens, the LM will generate ``filling token''. In this situation, we manually pad $N$ text tokens. If the text tokens run out of, the ``Turn of speech'' token \circled{T} will be added. In the streaming mode, we return generation results every $M$ tokens until the \circled{E} is detected.

\item \textbf{SFT, Non-Streaming:} In the SFT scenario, the LM is fine-tuned on a specific speaker, and the prompt text and speech are no longer needed. Thus, the initial sequence is very simple: ``\circled{\textbf{S}}, \textbf{text}, \circled{\textbf{T}}''. Starting from this, the text-speech LM can generate speech tokens autoregressively until \circled{T}.

\item \textbf{SFT, Streaming:} In the streaming mode of SFT, we start the speech generation from the following sequence: ``\circled{\textbf{S}}, \textbf{first\_N\_text}''. Then, the LM will generate $M$ speech tokens, and we manually pad the next $N$ text tokens. We repeat the above process until all text tokens run out of, and then \circled{\textbf{T}} is added. Note that this mode can also be adopted by the speech-to-speech multi-modal large language models to obtain an extremely low latency.
\end{itemize}

\subsection{Chunk-aware Flow Matching}
\begin{figure*}[t!]
	\centering
	\includegraphics[width=0.8\linewidth]{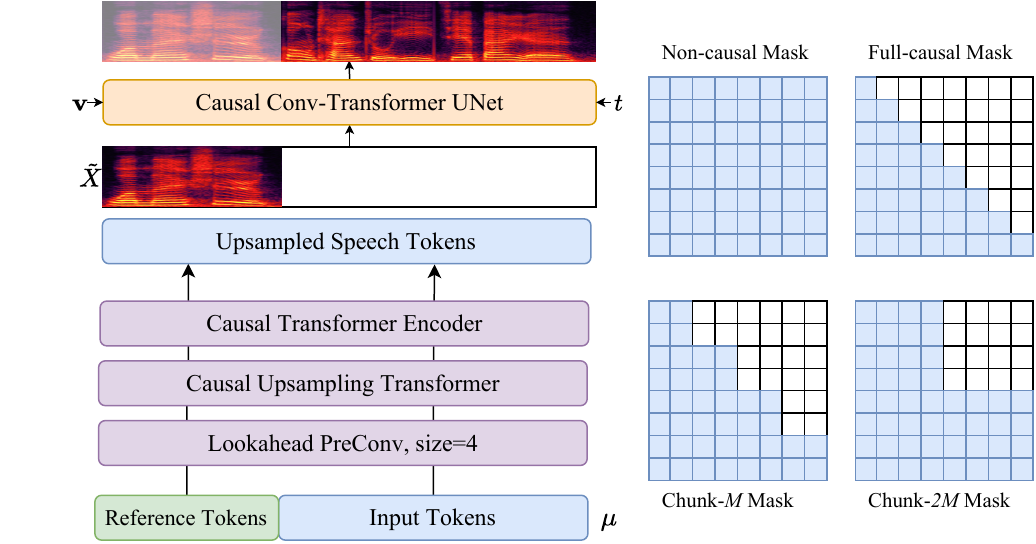}
	\caption{A diagram of the unified chunk-aware flow matching model for streaming and non-streaming synthesis in CosyVoice 2.}
	\label{fig:CFM}
\end{figure*}
In CosyVoice 2, we employ the Mel spectrogram as the acoustic feature with the frame rate of 50 Hz and the sampling rate of 24000. Due the frame-rate mismatch between speech tokens and Mel features, we up-sample the speech tokens with the ratio of two to match the frame rate of Mel spectrogram. Before the up-sampling operation, we add an additional look-ahead convolution layer to provide the future information for the following causal modules. The look-ahead layer is implemented by a right-padded 1-D convolution with the pad size of $P$ and the kernel size of $P+1$. After these, several chunk-aware causal Transformer blocks are followed to align the representation space of speech tokens to match acoustic features.

Subsequently, our goal is to further decode the speech tokens into the Mel spectrogram specified by the speaker embedding and reference speech.
To achieve this, we employ a conditional flow matching (CFM) model to sample the Mel spectrogram, given speech tokens, reference speech and speaker embedding as conditions. 
In the CFM model, the distribution of target Mel spectrogram is described by a probability density path from a prior distribution $p_0(X)$ and the data distribution $q(X)$. The probability density path can be defined by a time-dependent vector field.
For sampling efficiency, we employ the optimal-transport (OT) flow to match the vector field $\omega_t$, which is given by an ordinary differential equation (ODE):
\begin{align}
	\omega_t(\phi^{OT}_t(X_0,X_1)|X_1)&=X_1-X_0 \\
	\phi^{OT}_t(X_0,X_1)&=(1-t)X_0+tX_1 \\
	X_0&\sim p_0(X)=\mathcal{N}(0,I) \\
	X_1&\sim q(X)
\end{align}
A causal convolutional Transformer UNet is employed to learn the above ODE with the up-sampled token $\mu$, masked Mel spectrogram $\tilde{X}_1$, speaker embedding\footnote{https://github.com/alibaba-damo-academy/3D-Speaker/tree/main/egs/3dspeaker/sv-cam++} $\mathbf{v}$ and timestep $t$ as the conditions:
\begin{equation}
	\begin{aligned}
		\nu_t(\phi^{OT}_t(X_0,X_1)|\theta) &= \mathrm{UNet}_\theta\left(\phi^{OT}_t(X_0,X_1),t;\mathbf{v},\{\mu\}_{1:L},\tilde{X_1}\right) \\
	\end{aligned}
\end{equation}
At the training stage, the masked Mel spectrogram is obtained by randomly masking out 70\% to 100\% of the final frames in $X_1$. As for the inference, it is provided by the Mel spectrogram extracted from the reference speech. 
By minimizing the L1 loss between the predicted and ground-truth ODE, we can optimize the UNet parameters $\theta$ as follows:
\begin{equation}
	\theta = \arg\min_{\theta}\mathbb{E}_{p_0(X),q(X),t}{\left|\omega_t(\phi^{OT}_t(X_0,X_1))-\nu_t(\phi^{OT}_t(X_0,X_1)|\theta;\mu,\tilde{X}_1,\mathbf{v})\right|_1}
\end{equation}
At the training stage, the timestep follows a uniform distribution $U[0, 1]$. However, during the inference, we employ the cosine scheduler to offer more steps for the initial generation stage:
\begin{equation}
	t:=1-\cos\left(\frac{1}{2}t\pi\right)
\end{equation}
Besides, we also train the model on both conditional and non-conditional situations to enable the classifier-free guidance (CFG) \cite{ho2022classifier,nichol2021improved,le2024voicebox} at the inference stage:
\begin{equation}
	\begin{aligned}
		&\tilde{\nu}_t(\phi^{OT}_t(X_0,X_1)|\theta;\Psi)=(1+\beta)\cdot\nu_t(\phi^{OT}_t(X_0,X_1)|\theta;\Psi)-\beta \cdot \nu_t(\phi^{OT}_t(X_0,X_1)|\theta)
	\end{aligned}
\end{equation}
where $\Psi$ denotes the conditions $\{\mathbf{v},\mu,\tilde{X_1}\}$. The CFG strength $\beta$ and the number of flow estimation (NFE) are set to $0.7$ and $10$, respectively, according to the experimental results.

The current flow matching models always work on a offline mode, i.e., only all the speech tokens are generated, the Mel spectragram can be sampled, which is not friendly for the streaming synthesis.
To overcome this issue, we treat the multi-step flow estimation as a stacked deeper neural network, which repeats the UNet ten times. Thus, by making the unfolded neural network causal, we can apply it on the streaming synthesis. We construct four masks to satisfy different application situations:
\begin{itemize}[leftmargin=*]
\item \textbf{Non-causal Mask} is used for offline mode, which can achieve the best performance by attending all frames of conditions. Non-causal mask is suitable for the latency-insensitive situations.
\item \textbf{Full-causal Mask} is designed for scenarios required extremely low latency, in which only the past frames can be attended.
\item \textbf{Chunk-$M$ Mask} is a trade off between latency and performance, which can leverage the information of the past and $M$ future frames. This mask is more suitable for the first chunk of generation with low latency.
\item \textbf{Chunk-$2M$ Mask} can achieve a approximate performance of offline mode by sacrificing more latency, which can be used for the cascade generation chunk for better performance.
\end{itemize}
For each training case in a mini-batch, we randomly sample a mask from the above four masks under the uniform distribution. In this manner, one flow matching model can be compatible to different scenarios, lowering the deployment complexity. Another advantage of this chunk-aware training is that the masks with more context sever as a teacher for the ones with less context, benefiting from the implicit self-distillation scheme.

\subsection{Latency Analysis for Streaming Mode}
The first-package latency is an important metric for streaming synthesis models, which significantly affects the user experience especially in LLM-based voice chat applications, such as GPT-4o \cite{hurst2024gpt}.
In the context of TTS, the to-synthesize text is known in advance, and the latency comes from the aspects of speech token generation, Mel spectrogram reconstruction and waveform synthesis.
Thus, the first-package latency $L_{TTS}$ of CosyVoice 2 can be obtained as follows:
\begin{align}
L_{TTS} = M\cdot d_{lm} + M \cdot d_{fm} + M \cdot d_{voc}
\end{align}
where $d_{lm}$ denotes the computation time of LM to generate one speech token, $d_{fm}$ represents the computation time of Flow Matching model to generate the frames of Mel spectrogram for one speech token and $d_{voc}$ stands for the computation time of vocoder to synthesize waveforms corresponding to one speech token.
In the context of LLM-based voice chat, the length of first-package-required text should also be considered, and the first-package latency $L_{Chat}$ becomes as follows:
\begin{align}
	L_{Chat} \leq N\cdot d_{llm} + L_{TTS}
\end{align}
where $d_{llm}$ represents the computation time of a LLM to generate one text token. Note that, since the multi-character tokens are masked out in CosyVoice 2's text tokenizer, the text tokens used by text LLMs always encode longer raw text than those of CosyVoice 2. Thus, the the first-package latency $L_{Chat}$ must be lower than the summation of $N\cdot d_{llm}$ and $L_{TTS}$.


\subsection{Instructed Generation}

To enhance the controllability of CosyVoice 2, we integrated the instructed dataset into the base training set. We have collected 1500 hours of instructed training data, which includes both natural language instructions and fine-grained instructions, as outlined in Table \ref{tab:example_instruct}. For natural language instructions, we prepend a natural language description and a special end token, ``\texttt{<|endofprompt|>}'' before the to-synthesize input text. These descriptions cover aspects such as emotion, speaking rate, role-playing, and dialects. For fine-grained instructions, we insert vocal bursts between text tokens, using markers like ``\texttt{[laughter]}'' and ``\texttt{[breath]}''. Additionally, we apply vocal feature tags to phrases; for instance, ``\texttt{<strong>XXX</strong>}'' indicates emphasis on certain words, while ``\texttt{<laughter>XXX</laughter>}'' signifies speaking with laughter.

\begin{table}[h]
    \footnotesize
    \centering
    \scalebox{1.0}{
    \begin{tabularx}{\textwidth}{>{\raggedright\arraybackslash}X}
        \toprule
        \textit{Natural Language Instruction} \\
        \addlinespace
        \textbf{Emotion:} \begin{CJK}{UTF8}{gbsn}\footnotesize 高兴\end{CJK}(Happy), \begin{CJK}{UTF8}{gbsn}\footnotesize 悲伤\end{CJK}(Sad), \begin{CJK}{UTF8}{gbsn}\footnotesize 惊讶\end{CJK}(Surprised), \begin{CJK}{UTF8}{gbsn}\footnotesize 愤怒\end{CJK}(Angry), \begin{CJK}{UTF8}{gbsn}\footnotesize 恐惧\end{CJK}(Fearful), \begin{CJK}{UTF8}{gbsn}\footnotesize 厌恶\end{CJK}(Disgusted), \begin{CJK}{UTF8}{gbsn}\footnotesize 冷静\end{CJK}(Calm), \begin{CJK}{UTF8}{gbsn}\footnotesize 严肃\end{CJK}(Serious) \\
        \textbf{Speaking Rate:} \begin{CJK}{UTF8}{gbsn}\footnotesize 快速\end{CJK}(Fast), \begin{CJK}{UTF8}{gbsn}\footnotesize 非常快速\end{CJK}(Very Fast), \begin{CJK}{UTF8}{gbsn}\footnotesize 慢速\end{CJK}(Slow), \begin{CJK}{UTF8}{gbsn}\footnotesize 非常慢速\end{CJK}(Very Slow) \\
        \textbf{Dialect:} \begin{CJK}{UTF8}{gbsn}\footnotesize 粤语\end{CJK}, \begin{CJK}{UTF8}{gbsn}\footnotesize 四川话\end{CJK}, \begin{CJK}{UTF8}{gbsn}\footnotesize 上海话\end{CJK}, \begin{CJK}{UTF8}{gbsn}\footnotesize 郑州话\end{CJK}, \begin{CJK}{UTF8}{gbsn}\footnotesize 长沙话\end{CJK}, \begin{CJK}{UTF8}{gbsn}\footnotesize 天津话\end{CJK} \\
        \textbf{Role-playing:} \begin{CJK}{UTF8}{gbsn}\footnotesize 神秘\end{CJK}(Mysterious), \begin{CJK}{UTF8}{gbsn}\footnotesize 凶猛\end{CJK}(Fierce), \begin{CJK}{UTF8}{gbsn}\footnotesize 好奇\end{CJK}(Curious), \begin{CJK}{UTF8}{gbsn}\footnotesize 优雅\end{CJK}(Elegant), \begin{CJK}{UTF8}{gbsn}\footnotesize 孤独\end{CJK}(Lonely), \begin{CJK}{UTF8}{gbsn}\footnotesize 机器人\end{CJK}(Robot), \begin{CJK}{UTF8}{gbsn}\footnotesize小猪佩奇\end{CJK}(Peppa), etc. \\
        \midrule
        \textit{Fine-grained Instruction} \\
        \addlinespace
        \textbf{Vocal Bursts:} [laughter], [breath], etc. \\
        \textbf{Vocal Features:} $<$laughter$>$$<$/laughter$>$, $<$strong$><$/strong$>$ \\
        \midrule
        \textit{Examples} \\
        \addlinespace
        - \begin{CJK}{UTF8}{gbsn}\footnotesize 你能用高兴的情感说吗？\end{CJK}$<|$endofprompt$|>$\begin{CJK}{UTF8}{gbsn}\footnotesize 今天真是太开心了，马上要放假了！\end{CJK}I'm so happy, Spring Festival is coming! \\
        - Please speaking very fast.$<|$endofprompt$|>$Today is a happy day, full of laughter and joy. \\
        - \begin{CJK}{UTF8}{gbsn}\footnotesize 请问你能模仿粤语的口音吗？\end{CJK}$<|$endofprompt$|>$\begin{CJK}{UTF8}{gbsn}\footnotesize 多保重，早啲休息。\end{CJK}\\
        - \begin{CJK}{UTF8}{gbsn}\footnotesize 尝试一下以机器人的角色和我交流。\end{CJK}$<|$endofprompt$|>$\begin{CJK}{UTF8}{gbsn}\footnotesize 接收知识光波！\end{CJK}\\
        - \begin{CJK}{UTF8}{gbsn}\footnotesize [laughter]有时候，看着小孩子们的天真行为[laughter]，我们总会会心一笑。\end{CJK} \\
        - She pursued her dreams with $<$strong$>$enthusiasm$<$/strong$>$ and $<$strong$>$grit$<$/strong$>$. \\
        \bottomrule
    \end{tabularx}
    }
    \vspace{0.01cm}
    \caption{Examples of natural language instructions and fine-grained instructions.}
    \label{tab:example_instruct}
\end{table}

\subsection{Multi-Speaker Fine-tuning}
Fine-tuning the pre-trained model on specific speakers (SFT) can further improve the generation quality and speaker similarity. In this report, we introduce the multi-speaker fine-tuning (mSFT), in which the pretrained model is fine-tuned on multiply speakers simultaneously rather than a single speaker. This approach ensures comprehensive prosody and pronunciation coverage across multiple speakers and mitigates potential catastrophic forgetting from the pretrained models. To avoid timbre confusion between various speakers, we prepend speaker-prompt tags, ``\texttt{Speaker A<|endofprompt|>}'' to the input text for a specific speaker. If a training sample is not labeled to a speaker, a special tag, ``\texttt{unknown<|endofprompt|>}'', is utlized.
The learning rate is set to 1e-5 during the whole multi-speaker fine-tuning process.

\subsection{Reinforcement Learning for SFT}
Reinforcement learning is a commonly used method in the training of large language models, which can make the LM output align with human preference. 
In CosyVoice 2, we employ speaker similarity (SS) and recognition word error rate (WER) from the ASR system as the reward function to improve speaker similarity and pronunciation accuracy in the fine-tuning stage. We use WER and SS to distinguish preferred sample $x^w$ and rejected samples $x^l$ and optimize the TTS system with direct preference optimization (DPO) \cite{DBLP:conf/nips/RafailovSMMEF23} as follow:

\begin{equation}
    L_{DPO}(\pi_\theta; \pi_{\text{ref}}) = -\log \sigma(\beta \log \frac{\pi_\theta(\mu^w | y)}{\pi_{\text{ref}}(\mu^w | y)} - \beta \log \frac{\pi_\theta(\mu^l | y)}{\pi_{\text{ref}}(\mu^l | y)})
\end{equation}
where $\mu^w$ and $\mu^l$ are the speech token extracted from the preferred and rejected samples $x^w$ and $x^l$.



However, this method is time-consuming and computation-consuming as it should synthesis the audios through the TTS system repeatedly to obtain distinguishable preference and rejected samples. During training, four forward operations are needed for one training step.
To simplify the process, we recover the LM predicted token $\mu_i \in \{0, 1, \dots, (2K+1)^D-1\}$  into quantized low-rank representations $\bar{H}$, and directly use the ASR backend of the speech tokenizer to re-predict the input text. Then the predicted log posterior can be regarded as the ASR reward function to optimize the text-speech language model. During training, the ASR backend parameters are frozen.

\begin{equation}\label{func:codec}
        \bar{h}_{i,j} = \left\lfloor \frac{\mu_i}{(2K+1)^j} \right\rfloor \mod (2K+1)
\end{equation}

\begin{equation}
\begin{split}
    \hat{H} &= \mathrm{Proj}_{up}(\bar{H}) \\
    L_{ASR} &= -\log P(Y | \hat{H}; \theta_{ASR})
\end{split}
\end{equation}
where $Y$ is the input text, and $\bar{H}$ are the recovered speech low-rank representations. As the sample operation of the  $u_i \sim P(\mu_i|\mu_{1:i-1}, Y; \theta_{LM})$ still prevent us to optimize the model directly, we use the gumbel softmax sampling to make it differentiated and then optimize the $\theta_{LM}$ by the $\mathcal{L}_{ASR}$. 


\section{Experimental Settings}
\subsection{Training Data for Speech Tokenizer}

A 200,000-hour dataset is used to train the speech tokenizer with normalized transcriptions as labels. Detailed data information is listed in Table~\ref{tab:fsqdata}. The training data comes from three different resources: open source ASR datasets, internal industrial datasets and TTS generation datasets. Although we only used Chinese and English data when training the speech tokenizer, as shown in Table ~\ref{tab:fsqdata}, subsequent experiments revealed that the speech tokenizer had zero-shot capability for other languages. It can be also used for speech synthesis in languages such as Japanese and Korean.

\begin{table}[thb]
\centering
\scalebox{1.0}{
\begin{tabular}{lr}
\toprule
\textbf{Language} & \textbf{Duration (hours)} \\ \midrule
Chinese  & 110,884          \\
English  & 99,918           \\ \bottomrule
\end{tabular}
}
\vspace{0.1cm}
\caption{Details of training data for speech tokenizer.}
\label{tab:fsqdata}
\end{table}

\subsection{Training Data for CosyVoice 2}
CosyVoice 2 shares the same training data as its previous version \cite{cosyvoice}. 
We first collect the speech-only data with internal speech processing tools.
Subsequently, the Paraformer \cite{DBLP:conf/interspeech/GaoZ0Y22} and SenseVoice \cite{funaduiollm} are employed to generate pseudo text labels for Chinese and other languages, respectively. We also employ an internal force-alignment model to filter out low-quality data and enhances the accuracy of punctuation. Data details are provided in Table \ref{tab:cv_data}.

\begin{table}[thb]
\centering
\setlength\tabcolsep{10pt}
\scalebox{1.0}{
\begin{tabular}{lr}
\toprule
\textbf{Language}   & \textbf{Duration (hours)} \\ \midrule
Chinese & 130,000 \\
English & 30,000 \\
Japanese & 4,600 \\
Korean & 2,200 \\
\bottomrule
\end{tabular}
}
\vspace{0.1cm}
\caption{Details of training data for CosyVoice 2.}
\label{tab:cv_data}
\end{table}

\subsection{Evaluation Settings}
We evaluate our CosyVoice 2 on two test sets. The first one is constructed from the test-clean set of Librispeech corpus \cite{DBLP:conf/aaai/DuGSLLCWZ024}, denoted as \emph{test-clean}. This test set is used to evaluate CosyVoice 2 on a limited English domain. The Whisper-large V3 is used as the ASR model to evaluate the content consistency. As for the speaker similarity (SS), we employ the ERes2Net model \cite{eres2net} to extract speaker embeddings of prompt and generated utterances, and their raw cosine similarity is treated as the speaker similarity.
NMOS score \cite{DBLP:conf/icassp/ReddyGC22} is used to evaluate the objective quality.

The second evaluation is conducted on the SEED test sets \cite{DBLP:journals/corr/abs-2406-02430}, which is widely used to evaluate recent TTS models, covering various text domains and reference speeches. In this evaluation, about 2,000 Chinese and 1,000 English samples are selected from CommonVoice datasets, denoted as \emph{test-zh} and \emph{test-en}, respectively. In addition, about 400 hard test cases are also included to evaluate the robustness of TTS models on text repetition, tongue twister and other challenging synthesis cases, denoted as \emph{test-hard} in this report. The Paraformer is employed to recognize the synthesis results of \emph{test-zh} and \emph{test-hard}, while the Whisper-large V3 is adopted for \emph{test-en} to evaluate the content consistency. We adopt two speaker verification (SV) models to evaluate speaker similarity: WavLM-finetuned SV model and ERes2Net.

\subsection{Benchmark for Japanese and Korean}
We prepare two test sets, denoted as \emph{test-ja} and \emph{test-ko}, for the evaluation on Japanese and Korean speech synthesis. 
The \emph{test-ja} consists 1,000 samples extracted from the CommonVoice dataset, which are used to measure the model’s performance on various metrics, such as WER, SS, MOS.
Specifically, we randomly shuffle and pair the entire CommonVoice JA-test set as reference utterance and target utterance spoken.
Considering the wide range of utterances' text lengths of JA-test set, we randomly selected 1,000 pairs of reference-target utterances from the length range from 8 to 32 characters as our final test set.
For the \emph{test-ko}, we selected 1,000 speech samples with a WER of less than 5\% and no deletion or insertion errors, utilizing the Whisper-Large V3 \cite{DBLP:conf/icml/RadfordKXBMS23} as the ASR model. These samples were used as reference utterances for the Korean speech synthesis. For the input text, we randomly selected 1,000 text samples from the remaining data.
We have released the lists of prompt speeches, prompt transcriptions and input text from these two test sets are released to facilitate result reproduction. By providing this open-source data, we aim to establish a benchmark for evaluating Japanese and Korean TTS models.
The Whisper-large V3 is used as the ASR model for Japanese and Korean evaluations.

\section{Experimental Results}

\subsection{Evaluations on Speech Tokenizer}

An ideal speech tokenizer is supposed to effectively utilizes the codebook, preserves information at a high fidelity, and demonstrates speaker independence. In this part we evaluate our supervised speech tokenizer from four aspects: 1) Codebook utilization rate; 2) ASR error rate within the entire encoder; 3) Token visualization of different speakers; 4) Speaker identification training. Table~\ref{fsqres} shows the codebook utilization and ASR error rate. It turns out that the FSQ-based tokenizer fully utilizes the codebook and maintains more effective information from the aspect of ASR, indicating more semantic information maintained by FSQ.

\begin{table}[t]
\centering{%
\begin{tabular}{lcccccc}
\toprule
\multirow{2}{*}{\textbf{Method}} & \multicolumn{2}{c}{\textbf{Codebook}} & \multicolumn{4}{c}{\textbf{ASR Error Rate (\%)}}                                            \\ \cmidrule(r){2-3}  \cmidrule(r){4-7}
                  & \textbf{Size}     & \textbf{Util.}    & \textbf{C.V. EN} & \textbf{C.V. CN} & \textbf{Fluers EN} & \textbf{Fluers CN} \\ \midrule
\textbf{VQ}  & 4,096  & 963 (23\%)     & 18.26 & 11.56 & 7.65 & 5.03 \\
\textbf{FSQ} & 6,561 & 6,561 (100\%) & 10.67 & 7.29  & 6.58 & 4.43 \\ \bottomrule
\end{tabular}
}
\vspace{0.05cm}
\caption{The comparision of VQ and FSQ inside Sensevoice-large encoder. C.V. stands for the CommonVoice benchmarks.}
\label{fsqres}
\end{table}

\begin{figure*}[thb]
\centering
\includegraphics[width=\linewidth]{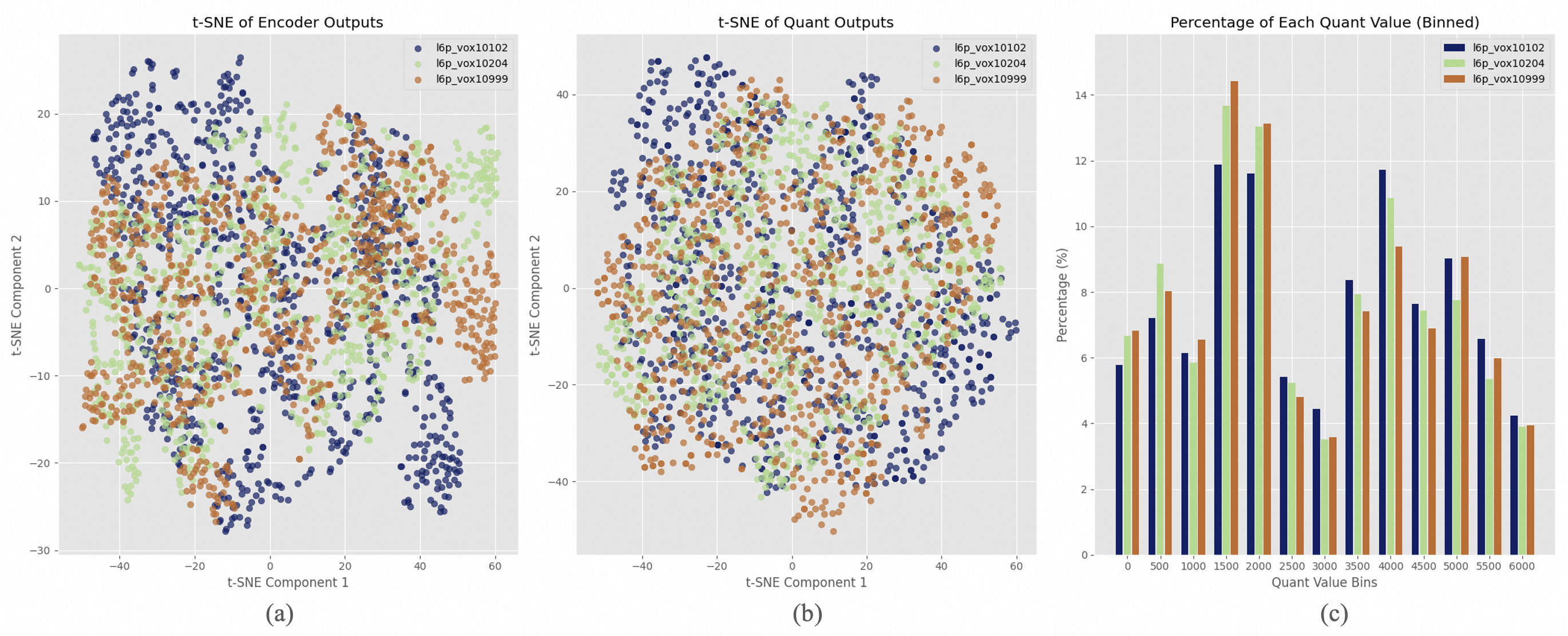}
\caption{The t-SNE visualization of speech representations before (a) and after (b) the quantization for three different speakers in Voxceb1 dataset. (c) shows the codebook utilization in terms of the token percentage on the speakers (500 tokens each bin).}
\label{fsqvis}
\end{figure*}

\begin{figure*}[thb]
\centering
\includegraphics[width=0.72\linewidth]{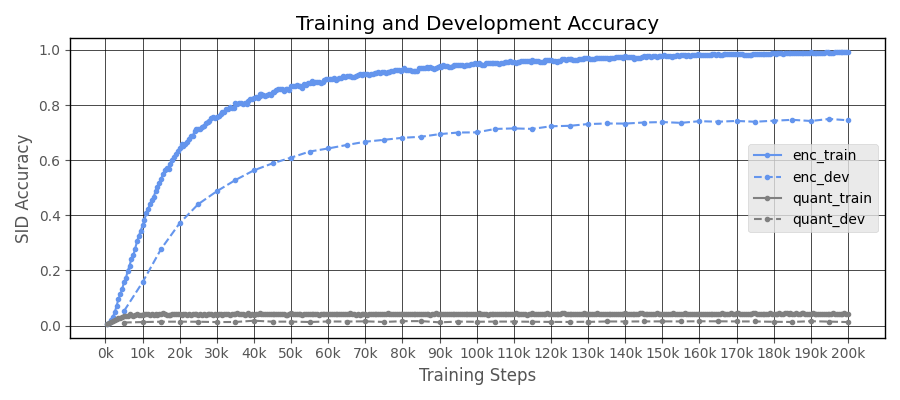}
\caption{The convergence curves of SID training with tokens before or after quantization.}
	\label{fsqtrain}
\end{figure*}

We further analyze the characteristics of FSQ through the t-SNE visualization. As an upstream model for TTS tasks, the tokenizer should strive to minimize the entanglement of speaker identity information with the speech signal. We selected 100 speech samples from each of the three speakers in the VoxCeleb1 dataset and visualized the corresponding tokens. As illustrated in Figures~\ref{fsqvis}(a) and (b), it is evident that before the quantization, Encoder$_1$'s outputs exhibit different distributions among different speakers. In contrast, the distributions of quantized representations are nearly indistinguishable. In addition, Figure~\ref{fsqvis}(c) also shows that the tokenizer fully utilizes the codebook. Subsequently, the S3prl toolkit \cite{yang2024large} is employed to further evaluate the speaker entanglement by performing speaker identification (SID) task. We use Sensevoice-large encoder with FSQ as an upstream feature extractor and train SID task with representations before or after the quantization. Figure~\ref{fsqtrain} shows the accuracy curves during the training. The SID layer with quantized tokens does not converge, which proves the decoupling function of the tokenizer on speaker information.

\subsection{Comparison Results with Baselines}
\begin{table*}[thb]
\centering
\setlength\tabcolsep{10pt}
\scalebox{1.0}{
    \begin{tabular}{l|ccc}
        \toprule
        \textbf{Model} & \textbf{WER (\%)} & \textbf{NMOS} & \textbf{SS} \\
        \midrule
        \textbf{Human} & 2.66 & 3.84 & 0.697 \\
        \midrule
        \textbf{ChatTTS} \cite{chatts} & 6.84 & 3.89 & - \\
        \textbf{GPT-SoVITs} \cite{gptsovits} & 5.13 & 3.93 & 0.405 \\
        \textbf{OpenVoice} \cite{DBLP:journals/corr/abs-2312-01479} & 3.47 & 3.87 & 0.299 \\
        \textbf{ParlerTTS} \cite{DBLP:journals/corr/abs-2402-01912} & 3.16 & 3.86 & - \\
        \textbf{EmotiVoice} \cite{EmotiVoice} & 3.14 & 3.93 & - \\
        \midrule
        \textbf{CosyVoice} \cite{cosyvoice} & 2.89 & 3.93 & 0.743 \\
        \textbf{CosyVoice 2} & 2.47 & \textbf{3.96} & 0.745 \\
        \textbf{CosyVoice 2-S} & \textbf{2.45} & 3.90 & \textbf{0.751} \\
        \bottomrule
\end{tabular}}
\caption{Content consistency (WER), speaker similarity (SS) and speech quality (NMOS) results on LibriSpeech test-clean subset of baselines and CosyVoice 2. Whisper-Large V3 is employed as the ASR model and punctuations are excluded before WER calculation.}
\label{tab:res-librispeech}
\end{table*}

We first evaluated our CosyVoice 2 models on a limited English text domain and compared it with several open-source models, such as ChatTTS \cite{chatts}, GPT-SoVITs \cite{gptsovits}, OpenVoice \cite{DBLP:journals/corr/abs-2312-01479}, ParlerTTS \cite{DBLP:journals/corr/abs-2402-01912}, EmotiVoice \cite{EmotiVoice}, and its predecessor CosyVoice \cite{cosyvoice}. The objective results are presented in Table \ref{tab:res-librispeech}, including content consistency (WER), speech quality (NMOS) and speaker similarity (SS). From the table, we can see that CosyVoice 2  achieves state-of-the-art performance on the Librispeech test-clean set, surpassing all baseline models acros all evaluation metrics. Notably, CosyVoice 2 even demonstrates higher content consistency, speech quality, and speaker similarity than human utterances, indicating its human-parity synthesis quality. 

\begin{table*}[thb]
	\centering
	\setlength\tabcolsep{6pt}
	\scalebox{0.80}{
		\begin{tabular}{lcccccc}
			\toprule
			\multirow{2}{*}{\textbf{Model}} & \multicolumn{2}{c}{\textbf{\emph{test-zh}}} & \multicolumn{2}{c}{\textbf{\emph{test-en}}} & \multicolumn{2}{c}{\textbf{\emph{test-hard}}} \\
			\cmidrule(r){2-3} \cmidrule(r){4-5} \cmidrule(r){6-7}
			& \textbf{CER (\%)~$\downarrow$} & \textbf{SS~$\uparrow$} & \textbf{WER (\%)~$\downarrow$} & \textbf{SS~$\uparrow$} & \textbf{WER (\%)~$\downarrow$} & \textbf{SS~$\uparrow$}  \\
			\midrule
			\textbf{Human} & 1.26 & 0.755~(0.775) & 2.14 & 0.734~(0.742)  & - & - \\
			\textbf{Vocoder Resyn.} & 1.27 & 0.720 & 2.17 & 0.700 & - & - \\
			\midrule
			\textbf{Seed-TTS}$^\dagger$~\cite{DBLP:journals/corr/abs-2406-02430} & 1.12 & 0.796 & 2.25 & 0.762  & 7.59 & 0.776 \\
			\textbf{FireRedTTS}~\cite{DBLP:journals/corr/abs-2409-03283} & 1.51 & 0.635~(0.653)  & 3.82 & 0.460~(0.526)  & 17.45 & 0.621~(0.639) \\
			\textbf{MaskGCT}~\cite{DBLP:journals/corr/abs-2409-00750} & 2.27 & 0.774~(0.752) & 2.62 & 0.714~(0.730)  & 10.27 & 0.748~(0.720) \\
			\textbf{E2 TTS (32 NFE)}$^\dagger$~\cite{DBLP:journals/corr/abs-2406-18009} & 1.97 & 0.730 & 2.19 & 0.710  & - & - \\
			\textbf{F5-TTS (32 NFE)}~\cite{DBLP:journals/corr/abs-2410-06885} & 1.56 & 0.741~(0.794) & 1.83 & 0.647~(0.742)  & 8.67 & 0.713~(0.762) \\
			\midrule
            \textbf{CosyVoice} \cite{cosyvoice} & 3.63 & 0.723~(0.775) & 4.29 & 0.609~(0.699)  & 11.75 & 0.709~(0.755) \\
			\textbf{CosyVoice 2} & 1.45 & 0.748~(0.806) & 2.57 & 0.652~(0.736)  &  6.83 & 0.724~(0.776) \\
			\textbf{CosyVoice 2-S} & 1.45 & 0.753~(0.812) & 2.38 & 0.654~(0.743)  & 8.08 & 0.732~(0.785) \\
			\bottomrule
	\end{tabular}}
	\caption{Results of CosyVoice 2 and recent TTS models on the SEED test sets. $\dagger$ denotes close-sourced models. For speaker similarity, the result in a bracket are measured by ERes2Net, while the results outside brackets are measured by WavLM-based models. }
	\label{tab:comapre}
\end{table*}

We also evaluated CosyVoice 2 on the commonly-used test sets: SEED \emph{test-zh}, \emph{test-en} and \emph{test-hard}, which include diverse input texts and reference speeches from various domains. The experimental results for CosyVoice 2 and the baseline models are presented in Table \ref{tab:comapre}. On the \emph{test-zh} set, CosyVoice 2 surpasses all open-sourced models in terms of CER and SS, falling short of the commercial model SEED-TTS by only a small margin. On the \emph{test-en} set, CosyVoice 2 ranks fourth and third in terms of WER and SS, respectively. This may result from the imbalance in the volume of training data between Chinese and English. We plan to explore data scaling in future work to enhance content consistency in English. On the \emph{test-hard} set, the offline CosyVoice 2 model achieves state-of-the-art performance across all compared baseline, demonstrating its robustness in challenging synthesis scenarios. Compared with human-generated speeches, CosyVoice 2 shows comparable content consistency and superior speaker similarity. Considering the recognition errors can also stem from the ASR model, it is reasonable to conclude that CosyVoice 2 achieves a human-parity synthesis capability. We also evaluated the streaming mode, denoted as ``CosyVoice 2-S'' in Table \ref{tab:res-librispeech} and \ref{tab:comapre}. For both evaluation settings, the streaming mode's performance is nearly lossless in typical test cases. Only in challenging cases is there a slight degradation in content consistency, highlighting the strength of our unified streaming/non-streaming framework. We found that the results of speaker similarity are not consistent on different SV models. This may indicate a new research topic on how to evaluate speaker similarity for TTS models automatically. Since different TTS models may use different SV models to extract speaker information, evaluating speaker similarity with the same SV model allows a more accurate evaluation on the utilization of speaker information. Therefore, we employ ERes2Net \footnote{https://github.com/modelscope/3D-Speaker} for evaluating speaker similarity in subsequent experiments.

\subsection{Modular Ablation Study}
We conducted a modular ablation study on the text-speech language model to assess the impacts of our modifications, including LLM initialization, removing speaker embedding, and utilizing FSQ. Table \ref{tab:modular} illustrates the step-by-step development of CosyVoice 2 from its predecessor. By replacing the randomly initialized language model with a pretrained LLM, we achieved relative improvements in content consistency of 18.46\% and 15.40\% on the \emph{test-zh} and \emph{test-hard} sets, respectively. Next, we removed the speaker embedding from the text-to-speech language model, which helps prevent information leakage and disturbances in in-context learning. This change resulted in a significant reduction in content errors while maintaining speaker similarity, indicating that content information is primarily modeled by the LM, and speaker information is mainly recovered by the flow matching model. Finally, by replacing VQ with FSQ, we achieved the CosyVoice 2 model, noting much higher content consistency and unchanged speaker similarity. By fully utilizing the codebook, FSQ captures more content information and context variation, leading to better alignment between text and speech tokens. Furthermore, we conducted a comparative experiment by incorporating pitch loss as a constraint during the training of the FSQ-based speech tokenizer. We found that this approach led to improved performance in downstream TTS tasks, as indicated in the last row of Table \ref{tab:modular}. In future versions of CosyVoice, we plan to carry out more detailed experiments and analyses.

\begin{table*}[thb]
\centering
\setlength\tabcolsep{6pt}
\scalebox{1.0}{
    \begin{tabular}{lcccccc}
        \toprule
        \multirow{2}{*}{\textbf{Model}} & \multicolumn{2}{c}{\textbf{\emph{test-zh}}} & \multicolumn{2}{c}{\textbf{\emph{test-en}}} & \multicolumn{2}{c}{\textbf{\emph{test-hard}}} \\
        \cmidrule(r){2-3} \cmidrule(r){4-5} \cmidrule(r){6-7}
        & \textbf{CER (\%)} & \textbf{SS} & \textbf{WER (\%)} & \textbf{SS} & \textbf{WER (\%)} & \textbf{SS}  \\
        \midrule
        \textbf{CosyVoice} & 3.63 & 0.775 & 4.29 & 0.699  & 11.75 & 0.755 \\
        \textbf{~~+ LLM init.} & 2.96 & 0.808 & 4.57 & 0.730  &  9.94 & 0.789 \\
        \textbf{~~~~+ Drop Spk Emb.} & 2.56 & 0.804 & 3.81 & 0.740  &  9.66 & 0.778 \\
        \textbf{~~~~~~+ FSQ (CosyVoice 2)} & 1.45 & 0.806 & 2.57 & 0.736  & 6.83 & 0.776 \\
           \textbf{~~~~~~~~+ Pitch Loss } & 1.19 & 0.802 & 2.40 & 0.728  & 6.29 & 0.769 \\
        \bottomrule
\end{tabular}}
\caption{Modular analysis on the modifications of text-speech language model.}
\label{tab:modular}
\end{table*}

We also conducted another modular analysis to evaluate the impact of streaming modules on the synthesis performance. Table \ref{tab:res-streaming} shows the results for content consistency and speaker similarity. We fount that the streaming LM has a minimal impact on typical cases from the \emph{test-zh} and \emph{test-en} sets, indicating the effectiveness of our unified training framework. The primary impact of the streaming LM is observed in challenging cases from the test-hard set, likely due to the loss of contextual information in streaming mode. Interestingly, the streaming flow matching model results in slightly higher speaker similarity compared to the offline mode. This may be due to the higher prompt-to-generation ratio of initial chunks in streaming mode, whereas the prompt-to-generation ratio in offline mode can be very low, with many padding tokens. The negative effect of the streaming flow matching model on content consistency is much less pronounced compared to streaming LMs, thanks to the semantic-acoustic decoupled modeling in CosyVoice 2.

\begin{table*}[thb]
\centering
\setlength\tabcolsep{6pt}
\scalebox{1.0}{
    \begin{tabular}{lcccccccc}
        \toprule
        \multirow{2}{*}{\textbf{Model}} & \multirow{2}{*}{\textbf{LM}} & \multirow{2}{*}{\textbf{FM}} & \multicolumn{2}{c}{\textbf{\emph{test-zh}}} & \multicolumn{2}{c}{\textbf{\emph{test-en}}} & \multicolumn{2}{c}{\textbf{\emph{test-hard}}}  \\
        \cmidrule(l){4-5} \cmidrule(l){6-7} \cmidrule(l){8-9}
         &  &  & \textbf{CER (\%)} & \textbf{SS} & \textbf{WER (\%)} & \textbf{SS} & \textbf{CER (\%)} & \textbf{SS} \\
        \midrule
        \textbf{M1} & Offline & Offline & 1.45 & 0.806 & 2.57 & 0.736 & 6.83 & 0.776 \\
        \textbf{M2} & Offline & Stream. & 1.46 & 0.811 & 2.60 & 0.743 & 7.12 & 0.788 \\
        \textbf{M3} & Stream. & Offline & 1.38 & 0.806 & 2.51 & 0.737 & 7.88 & 0.773 \\
        \textbf{M4} & Stream. & Stream. & 1.45 & 0.812 & 2.38 & 0.743 & 8.08 & 0.785 \\
        \bottomrule
\end{tabular}}
\caption{Modular analysis on the impact of streaming modules in CosyVoice 2. Chunk size is set to 15 for streaming modules.}
\label{tab:res-streaming}
\end{table*}

\subsection{Results on Japanese and Korean Benchmarks}
In addition to Chinese and English, CosyVoice 2 also supports Japanese and Korean. We evaluated the content consistency, speaker similarity and speech quality on our constructed Japanese and Korean test sets. As shown in Table \ref{tab:ja-ko}, CosyVoice 2 performs significantly better on Korean than on Japanese across all evaluation metrics. This discrepancy is primarily due to the overlap in the character set between Japanese and Chinese, which leads to Chinese pronunciations in Japanese contexts. In the future work, we plan to explore ways to enhance linguistic context for multilingual synthesis. Since Korean does not have character overlap with other languages, its speech synthesis achieves much better performance. Another issue is data imbalance. We believe that increasing the volume of training data could further improve synthesis performance for both Japanese and Korean.

\begin{table*}[thb]
\centering
\setlength\tabcolsep{6pt}
\scalebox{1.0}{
\begin{tabular}{lcccccc}
    \toprule
    \multirow{2}{*}{\textbf{Model}} & \multicolumn{3}{c}{\textbf{\emph{test-ja}}} & \multicolumn{3}{c}{\textbf{\emph{test-ko}}} \\
    \cmidrule(r){2-4} \cmidrule(r){5-7} 
    & \textbf{CER (\%)} & \textbf{SS} & \textbf{NMOS} & \textbf{CER (\%)} & \textbf{SS} & \textbf{NMOS}  \\
    \midrule
    \textbf{CosyVoice 2} & 18.79 & 0.630 & 3.42 & 7.98  & 0.707 & 3.73 \\
    \textbf{CosyVoice 2-S} & 21.41 & 0.629 & 3.35 & 9.06  & 0.714 & 3.60 \\
    \bottomrule
\end{tabular}}
\caption{The content consistency (CER), speaker similarity (SS), and speech quality (NMOS) of CosyVoice 2 and its streaming counterpart on the Japanese \emph{test-ja} and Korean \emph{test-ko} test sets.}
\label{tab:ja-ko}
\end{table*}

\subsection{Results on Instructed Generation}
To evaluate the performance of instructed generation, we have created a Chinese test set comprising 290 samples. This set includes 29 types of instructions, shown in Table \ref{tab:example_instruct}, each with 10 different input texts. We utilize five audio prompts and speaker embeddings from five speakers (three female and two male) as conditions for the flow matching model. Our testing is conducted in offline mode. We objectively evaluate content consistency (CER), speaker similarity (SS), and speech quality (NMOS). 
Subjectively, we assess the accuracy and naturalness of instruction using the Mean Opinion Score for Instruction (MOS-I), which ranges from 1 to 5. Each sample is assessed by 10 native Chinese speakers, with scores assigned in increments of 0.5. The evaluation criteria focus on whether the speech adheres to all specified instructions, such as emotional expression, speech rate adjustment, dialect usage, and role-playing. Fine-grained controls, including the insertion of laughter, speaking with laughter, breath control, and emphasis, are evaluated for naturalness and accuracy.
As illustrated in Table \ref{tab:res-instruct}, CosyVoice 2 exhibits superior content consistency (CER), speaker similarity (SS), and accuracy and naturalness in instruction control (MOS-I), while maintaining comparable speech quality to CosyVoice-Instruct. When input instructions are removed from CosyVoice 2, there is a notable decline in MOS-I; however, improvements are observed in content consistency (CER), speaker similarity (SS), and speech quality (NMOS). This indicates that instruction controllability is difficult to implicitly emerge from content text.

\begin{table*}[thb]
\centering
\setlength\tabcolsep{10pt}
\scalebox{1.0}{
    \begin{tabular}{l|cccc}
        \toprule
        \textbf{Model} & \textbf{CER (\%)} & \textbf{SS} & \textbf{NMOS} & \textbf{MOS-I}  \\
        \midrule
        \textbf{CosyVoice-Instruct} \cite{cosyvoice} & 1.72  & 0.797 & 3.94 & 3.09 \\
        \textbf{CosyVoice 2} & 1.52   & 0.804 & 3.94 & 4.06 \\
        \textbf{CosyVoice 2} w/o Instruction & 0.97  & 0.817 & 4.02 & 2.28 \\
        \bottomrule
\end{tabular}}
\caption{Evaluation results for content consistency (CER), speaker similarity (SS), speech quality (NMOS), and MOS-I (Instruction, assessing the accuracy and naturalness of instruction) on an in-house Chinese test set for CosyVoice-Instruct, CosyVoice 2, and CosyVoice 2 without instruction input. The Paraformer model is used as the ASR system, with punctuation marks excluded from the CER calculation. Dialect data is not included in the CER calculation because the Paraformer model cannot recognize Chinese dialect speech.}
\label{tab:res-instruct}
\end{table*}

\subsection{Results on Speaker Fine-tuned Models}

\begin{figure*}[thb]
\centering
\includegraphics[width=0.72\linewidth]{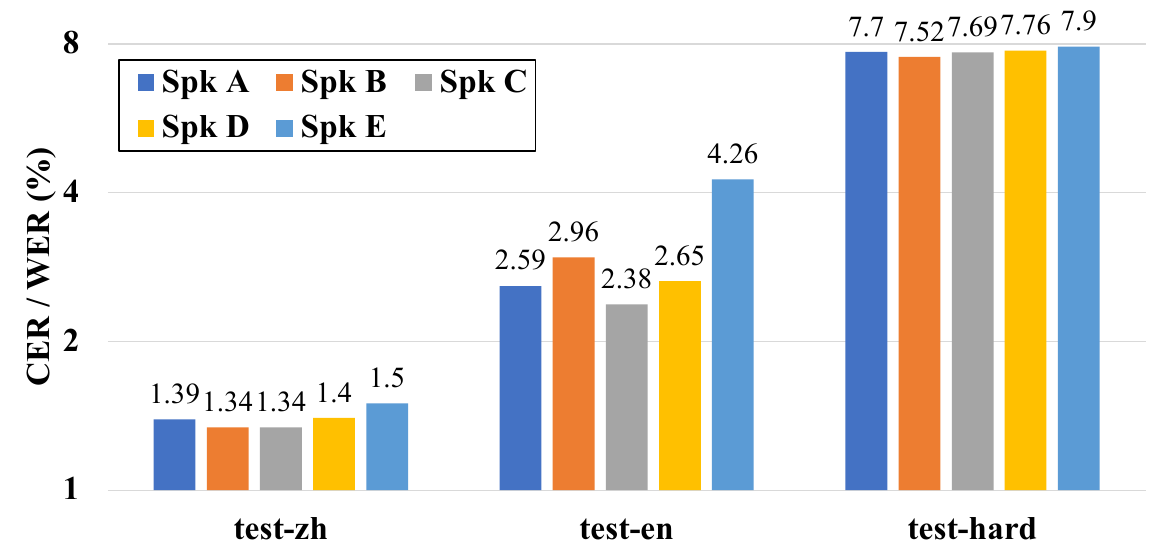}
\caption{Results of CosyVoice 2 SFT Models under the SEED evaluation settings. CER is used for test-zh and test-hard, while WER is used for test-en.}
	\label{fig:sft-fm}
\end{figure*}

During the fine-tuning phase, we employ unsupervised clustering on the speaker embeddings of the same speaker to ensure the stability of the speaker's timbre. We have demonstrated that a target speaker with as few as 400 audio recordings can achieve reasonably good speech synthesis performance, with only slight variations in objective metrics observed among different speakers, as shown in Figure \ref{fig:sft-fm}. Our experiments indicate that most speakers can inherit the zero-shot TTS model's robust contextual understanding and perception, thereby naturally expressing various moods and emotions in response to the input text.

\subsection{LM Fine-tuning with Reinforcement Learning}

Although the SFT can improve the performance on most speakers, the results of Spk E are still worse than the base model especially on English. Because Spk E has a more complex voice and faster speech speed. Additionally, only Chinese recordings are available for Spk E. So we apply the reinforcement learning on Spk E for further improvements. 
For DPO, we synthesis 10 thousand samples-pairs by the SFT models to change the preference biasing of the LM by the ASR and SS rewards. We also use the differentiable ASR rewards to optimize the LM parameters. 
After RL, we evaluate the model with content consistency (WER), speaker similarity (SS) and speech quality (NMOS) on the test set of Spk E and further evaluated the WER on the SeedTTS test sets to explore whether the model can maintain robustness to out-of-domain or cross-lingual input text. Results are shown in Table \ref{tab:res-sft}.

\begin{table*}[thb]
\centering
\setlength\tabcolsep{10pt}
\scalebox{1.0}{
    \begin{tabular}{l|ccc | c c c}
        \toprule
        \multirow{2}{*}{\textbf{Model}} &   \multicolumn{3}{|c|}{\textbf{Inhome Target Speaker}}  & \multicolumn{3}{c}{\textbf{SEED tests(\%)}}\\
         & \textbf{WER(\%)} & \textbf{NMOS} & \textbf{SS} & \textbf{zh}& \textbf{en} & \textbf{hard}\\
        \midrule
        Ground Truth        & 6.00 & 3.87 & 0.697 & 1.26 & 2.14 & -\\
        CosyVoice 2         & 5.34 & 3.91 & 0.721 & 1.45 & 2.57 & 6.83 \\
        \midrule
        CosyVoice 2-SFT     & 7.15 & 3.96 & 0.795 & 1.50 & 4.26 & 7.90\\
         \quad + $L_{ASR}$        & 6.79 & 3.96 & 0.795 & 1.29 & 3.53 & 7.30\\
         \quad + $L_{DPO}$      & 6.83 & 3.96 & 0.792 & 1.43 & 4.02 & 8.31\\
         \quad + $L_{ASR}$ +   $L_{DPO}$           & 6.64 & 3.97 & 0.796 & 1.25 & 3.17 & 6.66\\
        \bottomrule
\end{tabular}}
\caption{Content consistency (WER), speaker similarity (SS) and speech quality (NMOS) comparison for reinforcement learning models on Spk E.}
\label{tab:res-sft}
\end{table*}

Compared to the pre-trained base model, the SFT model shows higher speaker similarity and speech quality, however, the WER could be worse than the base model. 
We find that the audio synthesized by the base model always has a slower speed than the SFT and ground truth, which is more friendly to the ASR systems. 
For the target speaker dataset, both preference biasing and differentiable rewards can reduce the WER with little harmful effect on the other two metrics. 
But for the SEED test sets, the DPO based reinforcement only benefits the Chinese and English subset, while the hard samples will be worse. The reason could be that the hard samples contain many repeated words or phrases, they could be regarded as rejected samples during DPO training. 
However, the differentiable ASR reward will not suffer this problem, as it can directly optimize the TTS system by the ASR posterior.
This means that the differentiable ASR reward has a better generalization ability in the out-of-domain situations.
Finally, we can combine them with each other for further improvements. 


\section{Conclusion}
Building on the success of CosyVoice, this report presents CosyVoice 2, an improved streaming speech synthesis model that leverages large language models. By unifying streaming and non-streaming synthesis within a single framework, CosyVoice 2 achieves human-parity naturalness, minimal response latency, and virtually lossless synthesis quality in streaming mode. Key innovations include finite scalar quantization for full codebook utilization, a simplified text-to-speech language model architecture that incorporates pre-trained textual LLMs, and the development of a chunk-aware causal flow matching model to support diverse synthesis scenarios. Additionally, improvements in instructed TTS capacity allow for versatile and vivid speech generation with fine-grained control over emotion, accent, role style, and vocal bursts. Through systematic modifications and optimizations, CosyVoice 2 not only delivers superior synthesis quality but also loosens deployment requirements, making it suitable for both streaming and non-streaming applications. We believe that CosyVoice 2 represents a significant advancement in scalable, high-quality, and interactive text-to-speech synthesis.

\section{Limitations}
CosyVoice 2 has several limitations that need to be addressed. First, it supports only a limited number of languages. For languages with overlapping character sets, synthesis performance may degrade, presenting an open challenge for future research. Second, CosyVoice 2 cannot control acoustic characteristics, such as timbre, through textual instructions, which could be a fascinating area of exploration for role-playing applications. Additionally, CosyVoice does not perform well when tasked with singing.

\bibliographystyle{unsrt}
\bibliography{ref}

\appendix

\end{document}